\documentclass[pra,showpacs,nofootinbib,longbibliography,notitlepage]{revtex4-1}
\usepackage{hyperref}
\usepackage{amsmath}
\usepackage{amssymb}
\usepackage{amsthm}
\usepackage{graphics}
\usepackage{graphicx}
\usepackage{color}
\usepackage[usenames,dvipsnames,svgnames]{xcolor}

\long\def\ca#1\cb{}

\def\bra#1{\langle#1|}

\def\ket#1{|#1\rangle }

\def\Tr#1{\textrm{Tr}(#1)}

\def\AC{{\cal A}}
\def\BC{{\cal B}}
\def\CC{{\cal C}}

\def\HC{{\cal H}}

\def\KC{{\cal K}}
\def\LC{{\cal L}}
\def\MC{{\cal M}}
\def\NC{{\cal N}}

\def\PC{{\cal P}}
\def\QC{{\cal Q}}

\def\SC{{\cal S}}

\newtheorem{thm1}{Theorem}
\newtheorem{thm2}[thm1]{Theorem}

\newtheorem{lem1}[thm1]{Lemma}

\newtheorem{lem2}[thm1]{Lemma}
\newtheorem{lem3}[thm1]{Lemma}

\begin{document}
\title{Devising local protocols for multipartite quantum measurements}
\author{Scott M. Cohen}
\email{cohensm52@gmail.com}
\affiliation{Department of Physics, Portland State University, Portland OR 97201}

\begin{abstract}
We provide a method of designing protocols for implementing multipartite quantum measurements when the parties are restricted to local operations and classical communication (LOCC). For each finite integer number of rounds, $r$, the method succeeds in every case for which an $r$-round protocol exists for the measurement under consideration, and failure of the method has the immediate implication that the measurement under consideration cannot be implemented by LOCC no matter how many rounds of communication are allowed, including when the number of rounds is allowed to be infinite. It turns out that this method shows---often with relative ease---the impossibility by LOCC for a number of examples, including cases where this was not previously known, as well as the example that first demonstrated what has famously become known as nonlocality without entanglement.
\end{abstract}

\date{\today}

\maketitle
\section{Introduction}\label{sec1}
In recent years, a great deal of effort has been put forth toward understanding what is commonly referred to as \emph{local quantum operations and classical communication}, or LOCC. In this scenario, a global quantum system is shared amongst multiple spatially separated parties, each of those parties holding a subsystem in their respective laboratory. The parties aim to accomplish a given task, and the question is whether or not they can succeed without having to bring their subsystems together in a single laboratory. This is often an important question in practice, since bringing the subsystems together could be very costly in terms of time and other resources. It is also critical to our fundamental understanding of quantum information theory in general, and entanglement theory in particular, as LOCC is the class of operations that cannot increase entanglement (on average).

A ground-breaking study of LOCC was that given in \cite{Bennett9} of a specific state-discrimination problem, wherein the global system is prepared in one of a given set of mutually orthogonal quantum states, and the parties, initially ignorant of that choice, are to cooperate in an effort to determine which of those states was chosen, a task at which they are required to succeed every time. This is possible when the parties use a separable measurement \cite{Rains}, defined to be those for which each of the Kraus operators \cite{Kraus} representing the individual outcomes of the measurement is a tensor product, of the form $A\otimes B\otimes C\otimes\cdots$. However, if the parties are restricted to LOCC, they cannot perfectly discriminate the states \cite{Bennett9}. Since LOCC protocols always lead to product Kraus operators, every LOCC measurement is separable. As demonstrated (for the first time) by the study in \cite{Bennett9}, however, there exist separable measurements that cannot be implemented by LOCC.

Many other studies subsequently appeared in the literature, helping to illuminate the difference between separable measurements and LOCC, including several results showing a gap between what can be accomplished by these two classes of measurements for specific tasks such as quantum state discrimination \cite{Bennett9,IBM_CMP,NisetCerf,myLDPE,MaWehnerWinter,KKB,ChildsLeung,ChitHsieh1,ChitDuanHsieh,FuLeungManciska,ChitHsieh2} and the transformation of entangled states \cite{FortescueLo,Chitambar,ChitCuiLoPRL,ChitCuiLoPRA,WinterLeung}. In the latter case, it has been shown that this gap can be sizable \cite{ChitCuiLoPRL,ChitCuiLoPRA}.

In this paper, we consider quantum measurements, to be contrasted with quantum operations (the latter are also referred to as quantum channels). Both can be defined in terms of a set of Kraus operators, $\SC_K:=\{K_j\}$, but for a given measurement, $\SC_K$ is uniquely determined, whereas for a quantum channel, the choice of $\SC_K$ is far from unique. For a quantum measurement on input state $\rho$, outcome $j$ leaves the state as $\rho_j=K_j\rho K_j^\dag/p_j$ with the probability of that outcome given as $p_j=\Tr{K_j^\dag K_j\rho}$. The output of a quantum channel, on the other hand, is determined by the entire collection of Kraus operators as $\rho^\prime=\sum_jK_j\rho K_j^\dag$.

It is to be noted that LOCC measurements (and channels) are the only ones that can actually be implemented by spatially separated parties who lack the means to bring their systems together in one laboratory. Therefore it is significant that, while it is straightforward to recognize when a measurement is separable---one simply has to check if the Kraus operators are tensor products\footnote{We note that this is in stark contrast to the generally quite difficult problem of determining when a quantum \emph{channel} is separable, since unlike the case of a quantum measurement, there is not just a single set of Kraus operators that represent a quantum channel. Indeed, there is an entire continuous class of such sets for any given channel.}---it has been a challenge to identify methods of checking whether or not a measurement is LOCC. Studies of LOCC in the context of specific individual tasks, such as the one mentioned above by \cite{Bennett9}, have proved to be extremely enlightening. Nonetheless, it is of interest to find general ways of differentiating between separable measurements and LOCC, hopefully ones that provide insight into \emph{why} a given measurement can or cannot be implemented by LOCC. One such method is to check if there exists a local measurement that preserves orthogonality when the required task is to perfectly discriminate amongst a set of quantum states \cite{Walgate,WalgateHardy,myLDPE,myAlmostEvery,SingalOneWay}, since if there is no such measurement, then any measurement made by one of the parties will leave the set of states non-orthogonal, destroying any possibility for perfect discrimination. In this paper, we give an alternative general method that can be used to determine if any given measurement can be implemented by LOCC, regardless of the task that measurement is meant to accomplish.

Our method allows one to design an LOCC protocol to implement the desired measurement, whenever such a protocol exists. We have previously presented an alternative method of designing LOCC protocols \cite{mySEPvsLOCC,myMany}, but the one we give here is a very different approach. The previous method constructed a protocol starting from the end of that protocol and worked back toward its beginning, whereas the newer one we present below proceeds in the opposite direction, from beginning toward the end. The method presented below has the decided advantage that it can prove impossibility of a measurement by LOCC even when an infinite number of rounds of communication are allowed, whereas the previous method can only prove LOCC-impossibility when the number of rounds is finite. 

Before proceeding to a discussion of this alternative method, let us define the terminology and notation that will be used in the remainder of the paper. An LOCC protocol involves one party making a measurement, informing the other parties of her outcome, after which according to a pre-approved plan, the other parties know who is to measure next and what that measurement should be. This process---local measurement followed by communication---can generally proceed for as long as is necessary to achieve the parties' desired results. Since each measurement involves a number of possible outcomes, the entire process is commonly represented as a tree, the children of any given node representing the set of outcomes of the measurement made at that stage in the protocol. For any finite branch of the protocol, the leaf node at the end of that branch will represent a terminal outcome (leaf nodes being those that do not themselves have children), so must correspond to a desired outcome of the overall measurement to be implemented by the protocol. Consider the cumulative action of all parties up to a given node $n$ in the tree, represented by Kraus operator $K=A\otimes B\otimes\cdots$. We will label each such node by the positive operator $K^\dag K$, commonly referred to as a \emph{POVM element}, and we will say that node $n$ is `equal' to its label $K^\dag K$. For any such tree, the root node represents the situation present before any party has yet measured and will therefore always be equal to the identity operator $I=I_A\otimes I_B\otimes\cdots$ acting on the full multipartite Hilbert space $\HC$, where $I_\alpha$ is the identity operator on the local Hilbert space $\HC_\alpha$ for each party $\alpha$. We will refer to any tree labeled in this manner as an LOCC tree.

Our aim is to consider a given overall multipartite measurement and to design an LOCC protocol that implements that measurement, whenever such a protocol exists. The given measurement, which must be separable for the existence of an LOCC protocol to be possible, will be defined by a collection of Kraus operators, $\hat K_j=\hat A_j\otimes\hat B_j\otimes\cdots$, which is a more general definition than one given in terms of POVM elements. Nonetheless, it is only the associated POVM elements, $\hat\KC_j=\hat\AC_j\otimes\hat\BC_j\otimes\cdots$, where $\hat\AC_j=\hat A_j^\dag\hat A_j$, $\hat\BC_j=\hat B_j^\dag\hat B_j$, etc., that determine whether or not the measurement can be implemented by LOCC.\footnote{This is shown in Theorem~$2$ of \cite{myMany}.} We will always denote the operators defining the desired overall measurement with a carat over them, as we have done here. These operators will be referred to as \emph{the final outcomes} of the desired measurement.

An LOCC protocol can continue for any finite number of rounds, or it can have an infinite number of rounds, continuing indefinitely. We follow \cite{WinterLeung} in our classification of these protocols. When a measurement $\MC$ can be implemented by LOCC in a finite number of rounds, we say that $\MC\in$LOCC$_{\mathbb{N}}$. As discussed in \cite{WinterLeung}, the class of infinite-round protocols can be divided into two distinct sub-classes. One imagines a sequence of protocols whose limit is the infinite-round protocol under consideration. The first subclass, which is referred to as `LOCC' in \cite{WinterLeung}, involves sequences of protocols that more closely approach the desired measurement simply by adding more and more rounds of communication, but without changing what has been done in earlier rounds. In contrast, the second subclass, referred to as $\overline{\textrm{LOCC}}$ in \cite{WinterLeung}, includes sequences in which measurements made at the earlier rounds are changed from one protocol in the sequence to the next. As indicated by the notation, $\overline{\textrm{LOCC}}$ is the topological closure of LOCC, and it is known that the set of LOCC channels is not closed, the existence of a quantum channel in $\overline{\textrm{LOCC}}\backslash$LOCC having been demonstrated in \cite{ChitCuiLoPRL}, and then later in \cite{WinterLeung} for a two-qubit system (as far as we are aware, it is not presently known if the set of LOCC \emph{measurements} is closed). Following this classification, one may say that $\MC\in$LOCC or that $\MC\in\overline{\textrm{LOCC}}$, with the appropriate choice for the designated measurement. We note that the results of the present paper apply to the entire class LOCC, including those that involve an infinite number of rounds, but not to $\overline{\textrm{LOCC}}$. This is also true of arguments aimed at proving impossibility of perfect state discrimination by LOCC that are based on the fact that each measurement must preserve orthogonality of the states. Our approach is in some sense more general than those based on preserving orthogonality, because ours applies to all LOCC protocols, not just those aimed at the task of perfect state discrimination. However in order to apply our method, it is also the case that one needs to know exactly what the desired measurement is, a constraint that need not hold in applying the approach based on preserving orthogonality.

The remainder of the paper is organized as follows: In Sec.~\ref{sec2}, we make some observations about the structure of finite LOCC trees, and then extend these observations to infinite trees. In Sec.~\ref{sec3}, we show how to determine what `first' local measurements are allowed in initiating a protocol that must implement a given measurement, and then in Sec.~\ref{sec4}, we extend these ideas to any point in an LOCC protocol, showing how to determine which local measurements can possibly follow those local measurements that have gone before. In the Appendix, we provide a detailed example illustrating how to design an entire LOCC protocol for the chosen measurement. We will see that there exist cases where no local measurement is possible, either to initiate an LOCC protocol or to extend a protocol that has already proceeded through some number of rounds. Using this fact in Sec.~\ref{sec5}, we show how these ideas can be used to prove, often with relative ease, that certain measurements are impossible by LOCC. Finally, in Sec.~\ref{sec6}, we offer our conclusions.

\section{Observations on the structure of LOCC trees}\label{sec2}
In earlier work \cite{mySEPvsLOCC}, we labeled the nodes in an LOCC tree by the positive operator representing the cumulative action, to that point in the protocol, of the party who just measured to bring the protocol to that node in the tree.  Here, we will find it useful to instead label each node by the positive operator representing the cumulative action of all parties up to that point in the protocol. Thus, rather than a node being labeled by a \emph{local} operator, such as $\AC$, it will instead be labeled by the full multipartite operator, $\AC\otimes\BC\otimes\cdots$. In this case, following the very same type of argument used to obtain Eq.~(5) in \cite{mySEPvsLOCC} and Eq.~(3) in \cite{myMany}, it is easy to see that each node is now equal to the sum of all its child nodes. For example, if $K=A\otimes B\otimes C\otimes\cdots$ is the multipartite Kraus operator that has been implemented up to a given node, $n$, and party $B$, say, measures next with outcomes represented by Kraus operators $B_l$, then the overall Kraus operator implemented up to each of the children of node $n$ will be $K_l=A\otimes B_lB\otimes C\otimes\cdots$, respectively. With the completeness of the measurement represented by outcomes $B_l$, $\sum_lB_l^\dag B_l=I_B$, we see immediately that
\begin{align}\label{eqn21}
\sum_lK_l^\dag K_l = K^\dag K,
\end{align}
in accordance with our claim that each node is equal to the sum of all its children. Of course, this also means that each of those children is equal to the sum of its own children, implying that the original node is also equal to the sum of all of its grandchildren. Or more accurately, since some nodes---those that are terminal, or leaf, nodes---do not themselves have children, we see that each node in a finite-round LOCC tree is a sum of all of its descendant leaf nodes, a result that we state as the following lemma.
\begin{lem1}\label{lem1}
Each node $n$ in a finite-round LOCC tree is equal to the sum of the collection of all leaf nodes that are descendant from that node $n$.
\end{lem1}
\noindent Note that since leaf nodes are terminal, they must correspond to one of the final outcomes of the desired measurement. Therefore, each leaf node must be proportional (with positive constant of proportionality) to one of the set $\{\hat\AC_j\otimes\hat\BC_j\otimes\hat\CC_j\otimes\cdots\}$ representing the overall measurement implemented by the full LOCC protocol, and we therefore have the following result.
\begin{lem2}\label{lem2}
Suppose finite tree $\LC$ represents the finite-round LOCC protocol $\PC$, which implements overall measurement $\MC$. Then each node in $\LC$ is equal to a positive linear combination of the operators, $\{\hat\AC_j\otimes\hat\BC_j\otimes\hat\CC_j\otimes\cdots\}$, representing $\MC$. That is, if $\AC\otimes\BC\otimes\CC\otimes\cdots$ is the positive operator associated with any node in $\LC$, then
\begin{align}\label{eqn22}
\AC\otimes\BC\otimes\CC\otimes\cdots=\sum_jc_j\hat\AC_j\otimes\hat\BC_j\otimes\hat\CC_j\otimes\cdots,
\end{align}
with $c_j\ge0$.
\end{lem2}

Now let us consider the infinite-round case. Suppose we have a sequence of finite-round protocols, $\{\PC_m\}$, where (1) $\PC_{m+1}$ differs from $\PC_m$ only by the presence of one (or more) additional rounds added on at the end of the latter, (2) $\PC_m$ for any finite $m$ fails to exactly implement the desired overall measurement $\MC$, but (3) $\PC:=\lim_{m\to\infty}\PC_m$ does implement $\MC$ exactly, so that $\MC\in$LOCC$\backslash$LOCC$_{\mathbb{N}}$. Let $\MC$ be associated with the set of positive product operators,  $\{\hat\AC_j\otimes\hat\BC_j\otimes\hat\CC_j\otimes\cdots\}$. For each $\PC_m$, Lemma~\ref{lem1} holds, but leads to an error in implementation of $\MC$. Therefore, as $m$ increases and the number of rounds, $r$, becomes large enough, each node in the tree associated with that protocol in the sequence will be equal to a sum of the $\{\hat\AC_j\otimes\hat\BC_j\otimes\hat\CC_j\otimes\cdots\}$, apart from a small error. In the limit $m\to\infty$, each leaf must approach one of the operators $\{\hat\AC_j\otimes\hat\BC_j\otimes\hat\CC_j\otimes\cdots\}$, in order that this error approaches zero, which is necessary for $\PC$ to implement $\MC$. Hence, Lemma~\ref{lem2} may be generalized to the following important result, which will play a crucial role in the remainder of this paper.
\begin{lem3}\label{lem3}
Suppose tree $\LC$ represents the LOCC protocol $\PC$ (finite or infinite), which implements overall measurement $\MC$. Then each node in $\LC$ is equal to a positive linear combination of the operators, $\{\hat\AC_j\otimes\hat\BC_j\otimes\hat\CC_j\otimes\cdots\}$, representing $\MC$.
\end{lem3}

Let us now see what Lemma~\ref{lem3} tells us about the first measurement that can be made to initiate any LOCC protocol aimed at implementing a measurement $\MC$, whose associated POVM elements will be denoted as $\hat\AC_j\otimes\hat\BC_j\otimes\hat\CC_j\otimes\cdots$ throughout the remainder of this paper.

\section{The first measurement}\label{sec3}
Every LOCC protocol starts with one party making a first measurement. An obvious but, as we will soon see, very useful observation is that immediately after that first measurement has been made, all other parties have as yet done nothing. Therefore, the positive operator associated with a child of the root node of any LOCC tree must have a very special form. If, for example, party $A$ goes first obtaining measurement outcome $\AC$, then these positive operators are of the form, $\AC\otimes I_B\otimes I_C\otimes\cdots$, and as a direct consequence of Lemma~\ref{lem3}, we have that for each such outcome,
\begin{align}\label{eqn31}
\AC\otimes I_B\otimes I_C\otimes\cdots=\sum_jc_j\hat\AC_j\otimes\hat\BC_j\otimes\hat\CC_j\otimes\cdots,
\end{align}
for some set of positive coefficients, $c_j\ge0$.

Let $\SC_A$ and $\SC_{\bar A}$ be the subspaces spanned by operators $\{\hat\AC_j\}$ and $\{\hat\BC_j\otimes\hat\CC_j\otimes\cdots\}$, respectively. Consider bases of $\SC_A$ and $\SC_{\bar A}$, where the latter includes $I_{\bar A}:=I_B\otimes I_C\otimes\cdots$ as one member, and take the (tensor) product of these bases, denoting the resulting set of operators as $\{\QC_k\}$. Then, we can write
\begin{align}\label{eqn33}
\hat\AC_j\otimes\hat\BC_j\otimes\hat\CC_j\otimes\cdots=\sum_kq_{kj}\QC_k.
\end{align}
Find the respective dual bases to those bases just chosen for $\SC_A$ and $\SC_{\bar A}$,\footnote{If $\{X_j\}$ are the basis elements, then the dual basis is the unique set of elements, $\{Y_k\}$, which spans the same subspace as $\{X_j\}$ and such that $\Tr{Y_k^\dag X_j} = \delta_{jk}$.} omit the single operator in the latter dual basis that is \emph{not} orthogonal to $I_{\bar A}$, and take the (tensor) product of the first set of dual basis elements with the remaining operators in the second one, denoting the resulting set of operators as $\{\tilde\QC_l\}$. We have that
\begin{align}\label{eqn32}
\Tr{\tilde\QC_l^\dag\QC_k}=\delta_{kl},
\end{align}
and by design, each of the $\tilde\QC_l$ is orthogonal to $\AC\otimes I_{\bar A}$ for any $\AC$. Multiply both sides of Eq.~\eqref{eqn31} by any one of the $\tilde\QC_l^\dag$ and trace out all parties, which using Eq.~\eqref{eqn33} yields
\begin{align}\label{eqn34}
0=\sum_jc_j\sum_kq_{kj}\Tr{\tilde\QC_l^\dag\QC_k}=\sum_jq_{lj}c_j=:\mathbf{Q}\vec c,
\end{align}
where by Eqs.~\eqref{eqn33} and \eqref{eqn32}, matrix $\mathbf{Q}$ is seen to be formed from elements
\begin{align}\label{eqn35}
q_{lj}=\textrm{Tr}\left[\tilde\QC_l^\dag(\hat\AC_j\otimes\hat\BC_j\otimes\hat\CC_j\otimes\cdots)\right],
\end{align}
and the elements of vector $\vec c$ are given by the $c_j$.

We note that once the basis, $\{\QC_k\}$, is chosen, matrix $\mathbf{Q}$ is completely determined through Eq.~\eqref{eqn35}. Then, the set of solutions of Eq.~\eqref{eqn34} for coefficients $c_j$ determine, through Eq.~\eqref{eqn31}, the set of possible measurement outcomes in Alice's first measurement initiating the LOCC protocol. Thus, we have the following theorem.
\begin{thm1}\label{thm1}
If Alice measures first to initiate an LOCC protocol whose final outcomes are given by the set $\{\hat\AC_j\otimes\hat\BC_j\otimes\hat\CC_j\otimes\cdots\}$, the allowed outcomes of her initial measurement are determined by Eq.~\eqref{eqn31}, where coefficients $\{c_j\}$ must be the elements of a vector that lies in the nullspace of matrix $\mathbf{Q}$, whose matrix elements are given by Eq.~\eqref{eqn35}. [Recall that as noted below Eq.~\eqref{eqn33}, basis elements $\tilde\QC_l$ that are not orthogonal to $I_{\bar A}$ are not included in the construction of $\mathbf{Q}$.] Analogous conditions apply to any other party that wishes to measure first.
\end{thm1}
\noindent Note that while the choice of basis, $\{\QC_k\}$, is arbitrary, it is not difficult to see that the relevant nullspace is independent of this choice, as it must be.\footnote{To see this, note that any other basis has dual basis related to $\{\tilde\QC_k\}$ as $\tilde\QC_l^\prime=\sum_km_{lk}^\ast\tilde\QC_k$, where matrix $\mathbf{M}=\left(m_{lk}\right)$ is invertible. Then, from Eq.~\eqref{eqn35} we have $q_{lj}^\prime=\textrm{Tr}\left[\tilde\QC_l^{\prime\dag}(\hat\AC_j\otimes\hat\BC_j\otimes\hat\CC_j\otimes\cdots)\right]=\sum_km_{lk}\textrm{Tr}\left[\tilde\QC_k^\dag(\hat\AC_j\otimes\hat\BC_j\otimes\hat\CC_j\otimes\cdots)\right]=\sum_km_{lk}q_{kj}$, showing that $\mathbf Q^\prime = \mathbf{MQ}$ and by the invertibility of $\mathbf{M}$, $\mathbf Q^\prime$ and $\mathbf Q$ have identical nullspaces.}

As a very simple example, consider a separable measurement on two qubits consisting of POVM elements,
\begin{align}\label{eqn36}
\left\{\hat\AC_j\otimes\hat\BC_j\right\}=\left\{[0]\otimes[0]~,[0]\otimes[1]~,[1]\otimes[+]~,[1]\otimes[-]\right\},
\end{align}
where $[\psi]=\ket{\psi}\bra{\psi}$ and $\ket{\pm}=\left(\ket{0}\pm\ket{1}\right)/\sqrt{2}$. It is clear that the first party (Alice) can measure in the $\ket{0},\ket{1}$ basis followed by a measurement by the second party (Bob) in either the $\ket{0},\ket{1}$ or the $\ket{+},\ket{-}$ basis, depending on Alice's outcome. Nonetheless, let us analyze this example as an illustration of how Theorem~\ref{thm1} can be used. If Bob measures first, choose the bases of $\SC_A$ and $\SC_{B}$ to be $\{I_A,\sigma_z\}$ and $\{I_B,\sigma_z,\sigma_x\}$, respectively ($\sigma_z,\sigma_x$ are the usual Pauli operators). Since each of these bases is mutually orthogonal, we have that up to unimportant normalization factors, the $\{\tilde\QC_l\}$ are identical to the $\{\QC_k\}$ (except that some of the latter are omitted in the former, see the discussion above). Then omitting $I_A$ from the first basis, the relevant $\tilde\QC_l$ are $\sigma_z\otimes I_B,\sigma_z\otimes\sigma_z,\sigma_z\otimes\sigma_x$, and a simple calculation gives
\begin{align}\label{eqn39}
\mathbf{Q}&=\left[\begin{array}{cccc} 1&1&-1&-1\\
						1&-1&0&0\\
						0&0&-1&1\end{array}\right].
\end{align}
The nullspace of this matrix is the single ray, $\vec c=\left(c_1,c_1,c_1,c_1\right)^T$, and since $\sum_jc_1\AC_j\otimes\BC_j = c_1I_A\otimes I_B$, Bob's only allowed `outcome' is the identity operator, which means that there is no measurement that Bob can make at all. Thus, we see how our approach can immediately determine when a given party cannot initiate an LOCC protocol. While this is perhaps not entirely surprising for the present example (actually, it was proven previously in \cite{GroismanVaidman}), it turns out to be a very powerful, general observation, and in Sec.~\ref{sec5}, we discuss how these ideas can show that many measurements, including some that have been extensively studied, cannot be implemented by LOCC.

Can Alice measure first? In this case, we omit $I_B$ from the basis for $\SC_B$, so that the $\tilde\QC_l$ are now $I_A\otimes\sigma_z,I_A\otimes\sigma_x,\sigma_z\otimes\sigma_z,\sigma_z\otimes\sigma_x$, and we find
\begin{align}\label{eqn37}
\mathbf{Q}&=\left[\begin{array}{cccc} 1&-1&0&0\\
						0&0&1&-1\\
						1&-1&0&0\\
						0&0&1&-1\end{array}\right].
\end{align}
The nullspace of this matrix can be represented by all vectors of the form $\vec c=\left(c_1,c_1,c_3,c_3\right)^T$, and we see that
\begin{align}\label{eqn38}
\sum_jc_j\hat\AC_j\otimes\hat\BC_j = c_1\left([0]\otimes[0] + [0]\otimes[1]\right) + c_3\left([1]\otimes[+] + [1]\otimes[-]\right) = \left(c_1[0] + c_3[1]\right)\otimes I_B.
\end{align}
Hence, Alice's only allowed measurement outcomes are those that are diagonal in the $\ket{0},\ket{1}$ basis. In the following section, where we consider what subsequent measurements can be made following one or more previous measurements, we will see that in fact her only allowed outcomes are $[0]$ and $[1]$. This result will arise out of an application of Lemma~\ref{lem3}, from which we will find a very useful constraint on what a given party can do in a \emph{next} measurement, one that follows a given set of prior measurement outcomes.

\section{Additional measurements}\label{sec4}
Now that we have seen how to determine the possible first measurements, it is straightforward to generalize that approach to determine what measurements are allowed at any later stage of an LOCC protocol. That is, knowing what measurement outcomes have been obtained in preceding steps of a protocol, we can directly determine what measurements can be implemented next. Suppose that the parties have made some number of measurements, with their overall action to this point being represented by the operator $\AC\otimes\BC\otimes\CC\otimes\cdots$, and that Alice is to measure next. Her outcome will change $\AC$ to $\AC^\prime$, say, and Lemma~\ref{lem3} tells us that
\begin{align}\label{eqn41}
\AC^\prime\otimes\BC\otimes\CC\otimes\cdots=\sum_jc_j^\prime\hat\AC_j\otimes\hat\BC_j\otimes\hat\CC_j\otimes\cdots.
\end{align}
We again choose a basis of $\SC_A$ and $\SC_{\bar A}$, but for the latter, we want as many of the dual basis elements as possible to be orthogonal to $\bar\AC:=\BC\otimes\CC\otimes\cdots$ (rather than to $I_{\bar A}$, as was done before). So, choose $\bar\AC$ as one element of the $\SC_{\bar A}$ basis and then all but one of the elements in its dual basis will be orthogonal to $\bar\AC$, as desired. Then, omit that one element that is not orthogonal to $\bar\AC$, take the tensor product of the remaining elements with all elements in the basis dual to the basis chosen for $\SC_A$, and denote the resulting set as $\{\tilde\QC_l^\prime\}$. Multiply Eq.~\eqref{eqn41} by any one of the $\tilde\QC_l^{\prime\dag}$ and take the trace to obtain
\begin{align}\label{eqn42}
0=\sum_jc_j^\prime\sum_kq^\prime_{kj}\Tr{\tilde\QC_l^{\prime\dag}\QC_k^\prime}=\sum_jq^\prime_{lj}c_j^\prime=:\mathbf{Q^\prime}\vec c^{\,\prime},
\end{align}
where the elements of vector $\vec c^{\,\prime}$ are given by the $c^\prime_j$, and similarly to Eq.~\eqref{eqn35}, matrix $\mathbf{Q}^\prime$ is formed from elements
\begin{align}\label{eqn44}
q^\prime_{lj}=\textrm{Tr}\left[\tilde\QC_l^{\prime\dag}(\hat\AC_j\otimes\hat\BC_j\otimes\hat\CC_j\otimes\cdots)\right].
\end{align}
Thus, once again, we see that the allowed measurements are determined by the nullspace of a certain matrix, in this case $\mathbf Q^\prime$. As in the case of the first measurement, this is a linear condition on the $c_j^\prime$. However, it must be noted that since we needed to include $\bar\AC$ as a member of the basis chosen for $\SC_{\bar A}$, then $\mathbf Q^\prime$ depends on all previous measurement outcomes. Therefore, if one wishes to devise an entire LOCC protocol from the beginning, one no longer has the luxury of simply solving a set of linear equations. Rather, the constraints for the entire protocol are nonlinear, with the degree of nonlinearity increasing as the number of rounds increases. It is the case, however, that these constraints will be polynomial in the coefficients $c_j,c_k^\prime,$ and so on.

In the preceding section, we saw that for the measurement operators given in Eq.~\eqref{eqn36}, Alice can measure first with a measurement of $\AC=c_1[0] + c_3[1]$. We now consider Bob's next measurement and will see that this analysis further constrains the initial measurement by Alice. We choose a basis of $\SC_A$ as $\{\AC,\AC_\perp\}$, with $\AC_\perp = c_3[0] - c_1[1]$, and use the same basis for $\SC_{B}$ as before, $\{I_B,\sigma_z,\sigma_x\}$. Since these are each orthogonal bases, the dual bases will be identical to the original ones (again, up to unimportant normalization factors). Omitting $\AC$ from the $\SC_A$ basis and taking tensor products, we obtain $\{\tilde\QC^\prime_l\} = \{\AC_\perp\otimes I_B,\AC_\perp\otimes\sigma_z,\AC_\perp\otimes\sigma_x\}$, which from Eqs.~\eqref{eqn36} and \eqref{eqn44} then yields
\begin{align}\label{eqn45}
\mathbf{Q^\prime}&=\left[\begin{array}{cccc} c_3&c_3&-c_1&-c_1\\
						c_3&-c_3&0&0\\
						0&0&-c_1&c_1\end{array}\right].
\end{align}
If $c_1,c_3$ are both non-zero, the nullspace of $\mathbf Q^\prime$ is all vectors of the form $\vec c^{\,\prime} = \left(c_1,c_1,c_3,c_3\right)^T$, which leads to
\begin{align}\label{eqn46}
 \AC\otimes\BC^\prime = \sum_jc_j^\prime\hat\AC_j\otimes\hat\BC_j = \left(c_1[0] + c_3[1]\right)\otimes I_B.
\end{align}
Notice that Bob's local operator is identical to what it was before he ``measured" (in this case, $I_B$), implying that Bob cannot measure at all. On the other hand, if $c_3 = 0$ (that is, if Alice's initial outcome was $[0]$), then the nullspace of $\mathbf Q^\prime$ consists of vectors of the form $\vec c^{\,\prime} = \left(c_1^\prime,c_2^\prime,0,0\right)^T$, and then
\begin{align}\label{eqn47}
\AC\otimes\BC^\prime = \sum_jc_j^\prime\hat\AC_j\otimes\hat\BC_j = [0]\otimes\left(c_1^\prime[0] + c_2^\prime[1]\right),
\end{align}
and Bob can measure with any outcome diagonal in the $\ket{0},\ket{1}$ basis. Similarly, if $c_1 = 0$ (if Alice's initial outcome was $[1]$), then the nullspace of $\mathbf Q^\prime$ consists of vectors of the form $\vec c^{\,\prime} = \left(0,0,c_3^\prime,c_4^\prime\right)^T$, and then
\begin{align}\label{eqn47}
\AC\otimes\BC^\prime = \sum_jc_j^\prime\hat\AC_j\otimes\hat\BC_j = [1]\otimes\left(c_3^\prime[+] + c_4^\prime[-]\right),
\end{align}
and Bob can measure with any outcome diagonal in the $\ket{+},\ket{-}$ basis. Of course, in each of these cases, $c_1=0$ or $c_3=0$, the outcome of Alice's initial measurement is a rank-$1$ operator, which means that nothing she does subsequently can change this result. Therefore, the only logical thing for Bob to do in this second measurement is to measure $\{[0],[1]\}$ when Alice's outcome was $[0]$, or $\{[+],[-]\}$ when her outcome was $[1]$. Thus, we see that our approach has successfully determined a protocol that succeeds in implementing the desired measurement, and it has also demonstrated that this is effectively the only protocol that does so.

The utility of analyzing the nullspace of $\mathbf Q$ as a general approach to designing LOCC protocols should now be clear. In the Appendix, we use this `nullspace' approach to finding an LOCC protocol for an example that requires a good deal more effort than the one discussed above, demonstrating simplifications that may be useful for automating these analyses. It can, however, sometimes be instructive to take a more direct approach to this problem, even while continuing to use Lemma~\ref{lem3} and basing one's approach on the observation that when one party measures, the others do nothing. In the next section, we use both the nullspace analysis and also a more direct approach to show that these ideas are capable of demonstrating the LOCC-impossibility of various measurements, often in a quite straightforward way.

\section{Impossibility of a measurement by LOCC}\label{sec5}
Following the ideas presented in the preceding sections, it is possible to find all allowable LOCC protocols for a given overall final measurement. As illustrated in the example discussed above, one checks to see which parties can measure first to initiate a protocol, and for each party that can measure first, then at the next round one checks to see which parties can measure next, and so on until the procedure fails or the measurement is successfully implemented.\footnote{Notice that we can assume the party that just measured will not measure next. To see this, assume Alice just measured with outcome $\AC\otimes\BC\otimes\CC\otimes\cdots$, where the possible outcomes of her local measurement were found under the condition that $\BC\otimes\CC\otimes\cdots$ does not change when she measures. If she measures again next before any of the other parties measure, then the conditions determining her allowed outcomes have not changed. Therefore, her allowed local outcomes $\AC$ will also not have changed, so we may just as well omit her second (consecutive) measurement while still accounting for all possible ways in which the protocol can subsequently progress. We will therefore, without loss of generality, always assume that a party that just measured will not measure again until after a different party has measured in the interim.} To see how we get all possible protocols, recall that all allowed measurements at any given round and by any given party correspond to a vector in the nullspace of a matrix, $\mathbf{Q}$, defined by Eq.~\eqref{eqn35} or \eqref{eqn44}, and that nullspace is parametrized by a set of coefficients, say $c_j\ge0$. When we then consider the next measurement, we do so as a function of those coefficients parametrizing the immediately preceding measurement, and indeed, of those coefficients associated with all previous local measurements in the protocol up until the given round. In this way, we find all allowable next measurements following \emph{all} allowable preceding measurements, and thus we find all protocols for the given ordering of the parties. Hence, by considering all orderings of the parties,\footnote{For the first measurement one has $P$ parties that can measure first, and then there are $P-1$ parties that can possibly measure at each subsequent round. Therefore, there are on the order of $\left(P-1\right)^{\left(r-1\right)}$ orderings that need be considered for protocols with $r$ rounds. For $P\ge3$, the computational expense thus grows exponentially in $r$, though it does appear \cite{mySEPvsLOCC} that (at least) an exponential scaling in $r$ is unavoidable.} we find all possible protocols for the overall measurement.

We have seen in the preceding sections that our methods are capable of demonstrating that a given party is unable to measure next in an LOCC protocol that is required to exactly implement a given overall measurement. If none of the parties is able to measure next at some fixed point in our procedure (of designing LOCC protocols) before the desired measurement is fully implemented, then for that ordering of the parties, the desired overall measurement cannot be implemented. If this is true for all possible orderings of the parties, then this measurement cannot be exactly implemented by LOCC. Thus, our methods are capable of proving the LOCC-impossibility of a given measurement. While this conclusion can be reached at any stage of the process, it turns out that it is often sufficient to only investigate the very first measurement. That is, there are a wide array of separable measurements that can be proved impossible by LOCC via these methods by showing that no first measurement is possible by any of the parties, see below Theorem~\ref{thm2} and in Sec.~\ref{sec6}.

Before looking at specific cases, let us note that the nullspace of $\mathbf Q$, which we will henceforth denote as $\NC$, is always at least one-dimensional. For example, there always exists $\vec c$ such that $\sum_jc_j\hat\AC_j\otimes\hat\BC_j\otimes\hat\CC_j\otimes\cdots = I_A\otimes I_B\otimes I_C\otimes\cdots$, because by assumption, the overall measurement is complete. Therefore when considering the first local measurement to initiate an LOCC protocol, this $\vec c$ lies in $\NC$, which must therefore be at least one-dimensional as claimed. Similarly, when considering any later measurement from parent node $\AC\otimes\BC\otimes\CC\otimes\cdots$ to children $\AC_m^\prime\otimes\BC\otimes\CC\otimes\cdots$, say, then Lemma~\ref{lem3} tells us that there must exist $\vec c$ such that $\sum_jc_j\hat\AC_j\otimes\hat\BC_j\otimes\hat\CC_j\otimes\cdots = \AC\otimes\BC\otimes\CC\otimes\cdots$, or else the expression on the right-hand side of this equality could not have been that parent to begin with. Hence, once again, the claim is verified. If $\NC$ is exactly one-dimensional, then the only ray in $\NC$ lies along the given $\vec c$ that corresponds to that parent node, so the only children possible are in fact identical to their parent (up to a positive factor), which implies that the party under consideration cannot measure next. If this is true for every party, then there exists no next measurement for any one of the parties, and that parent node must then be terminal, a leaf node (and is then, in fact, not actually a parent). If that leaf is not one of the $\hat\AC_j\otimes\hat\BC_j\otimes\hat\CC_j\otimes\cdots$, then the procedure has failed to find a protocol that exactly implements the given overall measurement for the associated ordering of the parties' measurements, and if this is true for all possible orderings, we may then conclude the given overall measurement cannot be implemented by LOCC. Hence, we have the following theorem.
\begin{thm2}\label{thm2}
Let $\NC$ be the nullspace of matrix $\mathbf Q$, which is defined by the matrix elements given in Eq.~\eqref{eqn35} or \eqref{eqn44}, when considering a next measurement by party $\alpha$, which is to immediately follow a preceding measurement by party $\beta$. Then we have the following:
\begin{enumerate}
  \item If the dimension of $\NC$ is equal to unity, party $\alpha$ cannot measure next.
  \item\label{itm2} If the dimension of $\NC$ is equal to unity for all parties other than party $\beta$, then that branch must terminate at the given point.
  \item If the dimension of $\NC$ is equal to unity for all parties other than party $\beta$ and the (unique, normalized) vector defining $\NC$ does not correspond to one of the final measurement outcomes in the desired overall measurement $\MC$, then the associated ordering of the parties' measurements to that point in the procedure cannot implement $\MC$.
  \item If the dimension of $\NC$ is equal to unity for all parties other than party $\beta$ and the (unique, normalized) vector defining $\NC$ does not correspond to one of the final measurement outcomes in the desired overall measurement $\MC$, and if this eventuality occurs at some point in each of the protocols designed for all possible orderings of the parties, then $\MC\not\in$LOCC.
\end{enumerate}
\end{thm2}

Let us now illustrate this theorem through examples. In \cite{myExtViolate1}, we presented an infinite class of separable measurements and showed they could not be implemented by finite-round LOCC. Then, in \cite{myUSD}, we showed that each measurement in a subset of that class was the unique, optimal (and separable) measurement for unambiguously discriminating \cite{Ivanovic,Dieks,Peres} a corresponding set of states, and we therefore saw that this optimal measurement cannot be achieved by LOCC$_{\mathbb{N}}$. Here, we will take the simplest case from the latter subclass and use the ideas presented in the present paper to explicitly show this task cannot be achieved by LOCC, even when an infinite number of rounds are used. We note that it can be shown that the same conclusion holds for each member of this subclass; that is, none of these unique, optimal (and separable) measurements for unambiguous state discrimination can be achieved by LOCC, even with an infinite number of rounds.

The example we consider here involves two parties each holding a qubit system, and the set of states to be unambiguously discriminated has four members. As shown in \cite{myUSD}, the unique, optimal separable (and global) measurement for this task has rank-$1$ POVM elements that are proportional to projectors onto the following five states.
\begin{align}\label{eqn51}
\ket{\Psi_j}=\ket{\psi_j^{(1)}}\otimes\ket{\psi_j^{(2)}},~j=1,\ldots,5,
\end{align}
\noindent with
\begin{align}\label{eqn52}
\ket{\psi_j^{(\alpha)}}=\frac{1}{\sqrt{2}}\left(\ket{0}+e^{2\pi \textrm{i}jp_\alpha/5}\ket{1}\right),
\end{align}
where $p_1=1$ and $p_2=2$. It can be shown \cite{myExtViolate1} that
\begin{align}\label{eqn53}
I=\frac{2}{5}\sum_{j=1}^N\Psi_j,
\end{align}
with $\Psi_j=\ket{\Psi_j}\bra{\Psi_j}$. Defining $\omega=e^{\textrm{i}\theta}$ with $\theta = 2\pi /5$, and
\begin{align}\label{eqn54}
W_k=\frac{1}{2}\left[\begin{array}{cc} 1 & \omega^k\\
							 \omega^{-k} & 1
			\end{array}\right],
\end{align}
we have that 
\begin{align}\label{eqn55}
\Psi_j = W_j\otimes W_{2j}.
\end{align}
An orthogonal basis for the subspace $\SC_A$ spanned by the first party's local operators $W_k$ can be chosen to include the Pauli operators, as $\{I_{A(B)},\sigma_x,\sigma_y\}$ (these work also for $\SC_B$). Then if Alice measures first, we omit $I_B$ from the basis for $\SC_B$, choose the $\tilde\QC_l$ as $\{I_{A}\otimes\sigma_x,I_A\otimes\sigma_y,\sigma_x\otimes\sigma_x,\sigma_x\otimes\sigma_y,\sigma_y\otimes\sigma_x,\sigma_y\otimes\sigma_y\}$, and find from Eq.~\eqref{eqn35} that
\begin{align}\label{eqn510}
\mathbf{Q}&=\left[\begin{array}{ccccc} \cos2\theta&\cos\theta&\cos\theta&\cos2\theta&1\\
						\sin2\theta&-\sin\theta&\sin\theta&-\sin2\theta&0\\
						\cos\theta\cos2\theta&\cos\theta\cos2\theta&\cos\theta\cos2\theta&\cos\theta\cos2\theta&1\\
						\cos\theta\sin2\theta&-\sin\theta\cos2\theta&\sin\theta\cos2\theta&-\cos\theta\sin2\theta&0\\
						\sin\theta\cos2\theta&\cos\theta\sin2\theta&-\cos\theta\sin2\theta&-\sin\theta\cos2\theta&0\\
						\sin\theta\sin2\theta&-\sin\theta\sin2\theta&-\sin\theta\sin2\theta&\sin\theta\sin2\theta&0\\
			\end{array}\right].
\end{align}
On the other hand if Bob measures first, we omit $I_A$ from the basis for $\SC_A$, choose the $\tilde\QC_l$ as $\{\sigma_x\otimes I_B,\sigma_x\otimes\sigma_x,\sigma_x\otimes\sigma_y,\sigma_y\otimes I_B,\sigma_y\otimes\sigma_x,\sigma_y\otimes\sigma_y\}$, and find
\begin{align}\label{eqn511}
\mathbf{Q}&=\left[\begin{array}{ccccc} \cos\theta&\cos2\theta&\cos2\theta&\cos\theta&1\\
						\cos\theta\cos2\theta&\cos\theta\cos2\theta&\cos\theta\cos2\theta&\cos\theta\cos2\theta&1\\
						\cos\theta\sin2\theta&-\sin\theta\cos2\theta&\sin\theta\cos2\theta&-\cos\theta\sin2\theta&0\\
						\sin\theta&\sin2\theta&-\sin2\theta&-\sin\theta&0\\
						\sin\theta\cos2\theta&\cos\theta\sin2\theta&-\cos\theta\sin2\theta&-\sin\theta\cos2\theta&0\\
						\sin\theta\sin2\theta&-\sin\theta\sin2\theta&-\sin\theta\sin2\theta&\sin\theta\sin2\theta&0\\
			\end{array}\right].
\end{align}
In both Eqs.~\eqref{eqn510} and \eqref{eqn511}, the nullspace of $\mathbf Q$ is one-dimensional, given by all vectors proportional to $\left(1,1,1,1,1\right)^T$. Then, since $\sum_j\Psi_j\propto I_A\otimes I_B$ (or by Theorem~\ref{thm2}), we conclude that neither party can measure first, and therefore no LOCC protocol exists that exactly implements this measurement. 

Now we consider another example. Perhaps the most widely studied example of a separable measurement that cannot be implemented by LOCC was also the first such example of what has been called ``nonlocality without entanglement", discovered in \cite{Bennett9}. This problem involves the local distinguishability of a set of nine states on a $3\times3$ system. Here, we consider the more general case of what has been called the ``rotated domino states" \cite{ChildsLeung}, the projectors onto those nine states being
\begin{align}\label{eqn520}
\hat\AC_1\otimes\hat\BC_1 & = [1]_A\otimes[1]_B,\notag\\
\hat\AC_2\otimes\hat\BC_2 & = [0]_A\otimes\left(\cos\theta_2\ket{0}_B+\sin\theta_2\ket{1}_B\right)\left(\cos\theta_2\bra{0}_B+\sin\theta_2\bra{1}_B\right),\notag\\
\hat\AC_3\otimes\hat\BC_3 & = [0]_A\otimes\left(\sin\theta_2\ket{0}_B-\cos\theta_2\ket{1}_B\right)\left(\sin\theta_2\bra{0}_B-\cos\theta_2\bra{1}_B\right),\notag\\
\hat\AC_4\otimes\hat\BC_4 & = [2]_A\otimes\left(\cos\theta_4\ket{1}_B+\sin\theta_4\ket{2}_B\right)\left(\cos\theta_4\bra{1}_B+\sin\theta_4\bra{2}_B\right),\notag\\
\hat\AC_5\otimes\hat\BC_5 & = [2]_A\otimes\left(\sin\theta_4\ket{1}_B-\cos\theta_4\ket{2}_B\right)\left(\sin\theta_4\bra{1}_B-\cos\theta_4\bra{2}_B\right),\\
\hat\AC_6\otimes\hat\BC_6 & = \left(\cos\theta_6\ket{1}_A+\sin\theta_6\ket{2}_A\right)\left(\cos\theta_6\bra{1}_A+\sin\theta_6\bra{2}_A\right)\otimes[0]_B,\notag\\
\hat\AC_7\otimes\hat\BC_7 & = \left(\sin\theta_6\ket{1}_A-\cos\theta_6\ket{2}_A\right)\left(\sin\theta_6\bra{1}_A-\cos\theta_6\bra{2}_A\right)\otimes[0]_B,\notag\\
\hat\AC_8\otimes\hat\BC_8 & = \left(\cos\theta_8\ket{0}_A+\sin\theta_8\ket{1}_A\right)\left(\cos\theta_8\bra{0}_A+\sin\theta_8\bra{1}_A\right)\otimes[2]_B,\notag\\
\hat\AC_9\otimes\hat\BC_9 & = \left(\sin\theta_8\ket{0}_A-\cos\theta_8\ket{1}_A\right)\left(\sin\theta_8\bra{0}_A-\cos\theta_8\bra{1}_A\right)\otimes[2]_B,\notag
\end{align}
with $0<\theta_n\le\pi/4$. It has been shown that this measurement cannot be implemented by LOCC \cite{Bennett9,GroismanVaidman,myLDPE,ChildsLeung}. Nonetheless, let us offer one more proof of this fact. This can be done by finding the matrix $\mathbf Q$, defined in Eq.~\eqref{eqn35}, and showing that its nullspace is one-dimensional, but rather than writing down the $20\times9$ matrix $\mathbf Q$, we will take a somewhat more direct approach, which should offer additional insights into how/why this whole business works.

Let us consider the case that Alice measures first, so each of her initial outcomes must be equal to $\AC\otimes I_B= \sum_jc_j\hat\AC_j\otimes\hat\BC_j$, for some operator $\AC$ and non-negative coefficients $c_j$. Then, dividing the $9\times9$ matrix $\AC\otimes I_B$ into its nine $3\times3$ blocks, the diagonal blocks are found from Eqs.~\eqref{eqn520} to be respectively
\begin{align}\label{eqn521}
\left[\begin{array}{ccc} c_2\cos^2\theta_2+c_3\sin^2\theta_2 & \frac{1}{2}(c_2-c_3)\sin2\theta_2 & 0\\
							 \frac{1}{2}(c_2-c_3)\sin2\theta_2 & c_2\cos^2\theta_2+c_3\sin^2\theta_2 & 0\\
							0 & 0 & c_4\cos^2\theta_4+c_5\sin^2\theta_4
			\end{array}\right],
\end{align}
\begin{align}\label{eqn522}
\left[\begin{array}{ccc} c_8\cos^2\theta_8+c_9\sin^2\theta_8 & 0 & 0\\
							0 & c_1 & 0\\
							0 & 0 & c_4\cos^2\theta_4+c_5\sin^2\theta_4
			\end{array}\right],
\end{align}
and
\begin{align}\label{eqn523}
\left[\begin{array}{ccc} c_8\cos^2\theta_8+c_9\sin^2\theta_8 & 0 & 0\\
							 0 & c_6\cos^2\theta_6+c_7\sin^2\theta_6 & \frac{1}{2}(c_6-c_7)\sin2\theta_6\\
							0 & \frac{1}{2}(c_6-c_7)\sin2\theta_6 & c_6\cos^2\theta_6+c_7\sin^2\theta_6
			\end{array}\right].
\end{align}
There are also two pairs of non-zero off-diagonal blocks, each of the first pair being proportional to $(c_4-c_5)\sin2\theta_4[2]_B$ while the other pair are each proportional to $(c_8-c_9)\sin2\theta_8[0]_B$. Since the overall operator is $\AC\otimes I_B$, each block must be proportional to $I_B$. The only way the off-diagonal blocks can be proportional to $I_B$ is if the relevant proportionality constant is equal to zero. Therefore, we have that $c_4=c_5$ and $c_8=c_9$. The first diagonal block then tells us that $c_2=c_3=c_4$, the second diagonal block tells us that $c_8=c_1=c_4$, and in turn, the last diagonal block tells us that $c_6=c_7=c_8$. Hence, we see that the only possibility is $c_j=c$, the same for all $j$, and since $\sum_j\hat\AC_j\otimes\hat\BC_j=I_A\otimes I_B$, it must be that $\AC\propto I_A$, and we conclude that Alice cannot measure at all. In the same way using a similar analysis, one sees that Bob also cannot measure, and therefore we arrive at the conclusion that this set of states cannot be distinguished by LOCC.

Notice that in this section, the impossibility of LOCC has been demonstrated by showing that no party can measure first to initiate an LOCC protocol. These examples notwithstanding, LOCC-impossibility can clearly also be shown by demonstrating that no party can measure next at any later point in the design of such a protocol. Anecdotally, however, it appears to often (though certainly not always) be the case in known examples that no party can measure first to initiate a protocol, see Sec.~\ref{sec6} for further examples of this phenomenon.

\section{Conclusions}\label{sec6}
We have given a method of designing LOCC protocols to exactly implement quantum measurements, and for every finite integer $r$, the method provides an $r$-round protocol whenever one exists for the given measurement (in principle, the method will also provide an infinite-round protocol whenever one exists, but of course, it would take an infinite amount of time to find such a protocol). When the method fails to find an LOCC protocol, a direct conclusion is that the measurement cannot be implemented by LOCC, even when an infinite number of rounds of communication are allowed, see Theorem~\ref{thm2}. We have given examples where a protocol can indeed be obtained, see Secs.~\ref{sec3} and \ref{sec4}, and we have also given examples of measurements where this approach proves that the measurement cannot be implemented by LOCC, including an infinite class of examples \cite{myUSD} for which this conclusion had not been previously known, see Sec.~\ref{sec5}. While the examples in the main text involve only rank-$1$ operators in all cases, there is nothing about our method that restricts to such cases. Quite to the contrary, the method works regardless of the rank of the associated operators, as demonstrated by the class of examples given in the Appendix.

We note here that the method turns out to be extremely useful and can be employed to prove impossibility by LOCC in problems that go well beyond the few examples discussed above. Indeed, we have utilized a numerical implementation of these ideas to check the dimension of nullspace $\NC$ of matrix $\mathbf Q$, defined by Eq.~\eqref{eqn35}, to prove LOCC-impossibility for a wide range of cases that have appeared previously in the literature (and for which it has previously been shown that they cannot be implemented by LOCC). These cases include additional examples of nonlocality without entanglement, involving the perfect discrimination of certain complete, mutually orthogonal product bases, which can readily be obtained from associated unextendible product bases (UPB) \cite{IBM_PRL} (the projectors onto these basis states constitute a separable measurement that perfectly discriminates the given UPB). Examples to which we have applied this numerical approach include full orthogonal product bases derived from (i) the GenTiles1 UPB \cite{IBM_CMP}, which involves a $d\times d$ system with $d$ an even integer, where we have numerically demonstrated LOCC-impossibility for all even d with $4\le d\le36$ (beyond which we then run out of memory); (ii) GenTiles2  \cite{IBM_CMP}, which involves an $m\times n$ system with $n>3$, $m\ge3$, and $n\ge m$, and we have numerically shown that these states cannot be perfectly discriminated by LOCC for $3<m<10$, $m<n<100$, and for $11<m<36$, $m<n<36$ (beyond which we then ran out of patience); and (iii) the Niset-Cerf construction \cite{NisetCerf} of UPBs involving $P\ge3$ parties each with systems of dimension at least $P-1$, which we have considered cases where all $P$ systems are of the same dimension $d$ and shown numerically that these states cannot be perfectly discriminated by LOCC for cases when $P=3$ and $2\le d\le 11$ (after which we run out of memory), when P=4 and $2\le d\le6$ (and then run out of memory) and for $P=5$ and $d=4$ (beyond which we run out of memory). As mentioned in Sec.~\ref{sec4}, we have also proved that each member of the entire class of measurements given in \cite{myUSD}, each of which is optimal for unambiguous state discrimination, is impossible by LOCC, extending to infinite-round protocols the previously known result \cite{myUSD} that these measurements are impossible by LOCC$_\mathbb{N}$. We note that in all these cases, the impossibility by LOCC follows from the fact that no party can measure first to initiate an LOCC protocol.

\noindent\textit{Acknowledgments} --- We thank Li Yu for a helpful discussion. This work has been supported in part by the National Science Foundation through Grant No. 1205931.

\appendix
\section{Designing an LOCC measurement requiring more than one round of communication}
Consider a class of measurements having seven product operators, for which the local positive operators are related as
\begin{align}\label{eqnA1}
	\hat \BC_1&=2\hat \BC_2=3\hat \BC_3\notag\\
	\hat \BC_5&=\frac{1}{2}\left(I_B-2\hat \BC_1-\hat \BC_4\right)\notag\\
	\hat \BC_6&=\hat \BC_1+\hat \BC_4\notag\\
	\hat\BC_7&=I_B-\hat \BC_1-\hat \BC_4\\
	\hat \AC_4&=\frac{1}{2}\left(\hat \AC_1+\hat \AC_2\right)\notag\\
	\hat \AC_5&=\frac{1}{3}\left(\hat \AC_1+\hat \AC_3\right)\notag\\
	\hat \AC_6&=I_A-\hat \AC_1-\hat \AC_2\notag\\
	\hat \AC_7&=I_A-\hat \AC_1-\hat \AC_3\notag.
\end{align}
An LOCC protocol for this class of measurements was obtained as Example $5$ in \cite{mySEPvsLOCC} by an entirely different method than that given in the present paper. Here, for simplicity, we will also assume that $\SC_A=\{\hat\AC_0,\hat\AC_1,\hat\AC_2,\hat\AC_3\}$ and $\SC_B=\{\hat\BC_0,\hat\BC_1,\hat\BC_4\}$ is each a linearly independent set of operators, where $\hat\AC_0:=I_A$ and $\hat\BC_0:=I_B$. Let us first note that any linear combination of our seven product operators can be written as
\begin{align}\label{eqnA2}
\sum_{j=1}^7c_j\hat\AC_j\otimes\hat\BC_j	&= c_1\hat\AC_1\otimes\BC_1 + \frac{1}{2}c_2\hat\AC_2\otimes\hat\BC_1 + \frac{1}{3}c_3\hat\AC_3\otimes\hat\BC_1 + \frac{1}{2}c_4\left(\hat\AC_1 + \hat\AC_2\right)\otimes\hat\BC_4 \notag\\
				& + \frac{1}{6}c_5\left(\hat\AC_1 + \hat\AC_3\right)\otimes\left(I_B - 2\hat\BC_1 - \hat\BC_4\right) + c_6\left(I_A - \hat\AC_1 - \hat\AC_2\right)\otimes\left(\hat\BC_1 + \hat\BC_4\right) \notag\\
				&+ c_7\left(I_A - \hat\AC_1 - \hat\AC_3\right)\otimes\left(I_B - \hat\BC_1 - \hat\BC_4\right)\\
				&= \left[c_7I_A + \left(\frac{1}{6}c_5 - c_7\right)\left(\hat\AC_1 + \hat\AC_3\right)\right]\otimes I_B\notag\\
				& + \left[\left(c_6 - c_7\right) I_A + \left(c_1 - \frac{1}{3}c_5 - c_6 + c_7\right)\hat\AC_1 + \left(\frac{1}{2}c_2 - c_6\right)\hat\AC_2 + \left(\frac{1}{3}c_3 - \frac{1}{3}c_5 + c_7\right)\hat\AC_3\right]\otimes\hat\BC_1\notag\\
				& + \left[\left(c_6 - c_7\right)I_A + \left(\frac{1}{2}c_4 - \frac{1}{6}c_5 - c_6 + c_7\right)\hat\AC_1 + \left(\frac{1}{2}c_4 - c_6\right)\hat\AC_2 + \left(c_7 - \frac{1}{6}c_5\right)\hat\AC_3\right]\otimes\hat\BC_4.\notag
\end{align}
For our purposes, we wish to consider this sum as being equal to a given node in an LOCC tree, so it must be equal to a product operator, say $\AC\otimes\BC$. Now, if Alice can measure first, each node corresponding to the outcomes of her first measurement will have $\BC\propto I_B$. Since $\{I_B,\hat\BC_1,\hat\BC_4\}$ is linearly independent, this requires that the coefficients of $\hat\BC_1$ and $\hat\BC_4$ in the preceding equation must vanish. Then also using the linear independence of $\{I_A,\hat\AC_1,\hat\AC_2,\hat\AC_3\}$, it must be that $c_6=c_7$ and $c_5=3c_1=3c_4=6c_7$ so $c_1=c_4$, $c_2=2c_6=c_4$, and $c_3=3c_7$. Since $c_5=6c_7$, we have from the $I_B$ part of Eq.~\eqref{eqnA2} that $\AC=c_7I_A$, implying that Alice cannot measure first.

Let us now utilize the approach outlined in the main text. We first define dual bases for $\SC_A,\SC_B$ as $\tilde\SC_A=\{\tilde\AC_0,\tilde\AC_1,\tilde\AC_2,\tilde\AC_3\}$ and $\tilde\SC_B=\{\tilde\BC_0,\tilde\BC_1,\tilde\BC_4\}$, respectively, where $\Tr{\tilde\AC_i^\dag\hat\AC_j}=\delta_{ij}$ and $\Tr{\tilde\BC_i^\dag\hat\BC_j}=\delta_{ij}$. Form the set, $\tilde\QC_l$, from the tensor products of all operators in $\tilde\SC_A$ with all those in $\tilde\SC_B$, and define $\mathbf{Q}^{all}$ to have matrix elements $q^{all}_{lj}=\Tr{\tilde\QC_l^\dag\hat\AC_j\otimes\hat\BC_j}$. Then, we find
\begin{align}\label{eqnA3}
\mathbf{Q}^{all}=\left[\begin{array}{ccccccc}0         &0         &0         &0         &0         &0    &1\\
        0         &0         &0         &0         &0    &1   &-1\\
        0         &0         &0         &0         &0    &1  & -1\\
         0         &0         &0         &0    &1/6         &0   &-1\\
	1         &0         &0         &0   &-1/3   &-1    &1\\
        0         &0         &0    	&1/2   &-1/6   &-1    &1\\
        0         &0         &0         &0         &0         &0         &0\\
        0    	&1/2     &0         &0         &0   &-1         &0\\
        0         &0         &0    	&1/2         &0   &-1         &0\\
        0         &0         &0         &0    &1/6         &0   &-1\\
        0         &0        &1/3         &0   &-1/3        &0    &1\\
        0         &0        &0         &0   &-1/6         &0    &1
\end{array}\right]
\end{align}
The rows of this matrix in top-to-bottom order correspond to operators $\tilde\QC_l$ respectively as $\tilde\AC_0\otimes \tilde\BC_0,\tilde\AC_0\otimes \tilde\BC_1,\tilde\AC_0\otimes\tilde\BC_4,\tilde\AC_1\otimes \tilde\BC_0,\tilde\AC_1\otimes\tilde\BC_1,\tilde\AC_1\otimes\tilde\BC_4,\tilde\AC_2\otimes \tilde\BC_0,\tilde\AC_2\otimes\tilde\BC_1,\tilde\AC_2\otimes\tilde\BC_4,\tilde\AC_3\otimes \tilde\BC_0,\tilde\AC_3\otimes\tilde\BC_1,\tilde\AC_3\otimes\tilde\BC_4$.

Note that this matrix has more rows than would appear in any of the matrices $\mathbf{Q}$ that will appear when using the method described in the main text. However, we can save computational expense by using $\mathbf{Q}^{all}$ to obtain all the $\mathbf{Q}$ matrices needed to analyze every measurement in a complete LOCC protocol, from the very first one to the last. For the first measurement by either party, we just need to discard those rows in $\mathbf{Q}^{all}$ that involve the other party's identity operator in the operators, $\tilde\QC_l$. For Alice going first, this means we discard rows $1,4,7,10$, while for Bob going first, discard rows $1,2,3$. Equivalently, we can multiply $\mathbf{Q}^{all}$ by a diagonal matrix that has zeros in the diagonal element corresponding to those rows we wish to discard, and ones in all the other entries along the diagonal. This is because a row of zeros in $\mathbf{Q}$ is irrelevant for our purposes (clearly, we can choose to discard row $7$ in $\mathbf{Q}^{all}$ from the outset). For subsequent measurements, use of $\mathbf{Q}^{all}$ is a bit more involved, and we will see it is generally necessary to multiply $\mathbf{Q}^{all}$ by a non-diagonal matrix. In general, it is necessary to take linear combinations of the rows of $\mathbf{Q}^{all}$ in a way that uses only those $\tilde\QC_l$ that are orthogonal to the preceding measurement outcome of the parties (collectively denoted $\bar A$ in the discussion of the main text) that are not about to measure. For the present example, if Alice is to measure after Bob just obtained outcome $\BC=q\BC_0+r\BC_1+r\BC_4$, then we want Bob's part of the $\tilde\QC_l$ to be orthogonal to this operator $\BC$. This can be done in various ways, but one way, which we will utilize below, is to choose Bob's part of the $\tilde\QC_l$ to be either $r\tilde\BC_0+\omega q\tilde\BC_1+\omega^2q\tilde\BC_4$ or $r\tilde\BC_0+\omega^2q\tilde\BC_1+\omega q\tilde\BC_4$ [$\omega$ is defined below just before Eq.~\eqref{eqnA501}], which for the first choice corresponds to adding $r$ times the first row of $\mathbf{Q}^{all}$ to $\omega q$ times the second row and $\omega^2q$ times the third row, and making similar combinations of the other sets of three consecutive rows in $\mathbf{Q}^{all}$. See the discussion below for explicit demonstration of how this works.

We have already seen that Alice cannot measure first, so now let us consider if Bob can measure first. Discarding the first three rows of $\mathbf{Q}^{all}$, we find that the nullspace of what remains is spanned by the two orthogonal vectors $\left(2,2,3,2,6,1,1\right)^T$, which corresponds to $I_A\otimes I_B$, and $\left(37,96,-33,96,-66,48,-11\right)^T$, where superscript $T$ indicates the transpose operation. Since this nullspace is two-dimensional, Bob can indeed measure first, with outcomes of the form
\begin{align}\label{eqnA301}
\vec c^{\,T} = \left(c_1,c_2,c_3,c_4,c_5,c_6,c_7\right) = \left(c_6+c_7,2c_6,3c_7,2c_6,6c_7,c_6,c_7\right).
\end{align}
This corresponds to outcomes $I_A\otimes\BC=\sum_jc_j\hat\AC_j\otimes\hat\BC_j$, with
\begin{align}\label{eqnA5}
	\BC = c_7I_B + \left(c_6 - c_7\right)\left(\hat\BC_1 + \hat\BC_4\right) = c_6\hat\BC_6 + c_7\hat\BC_7.
\end{align}
[This is perhaps more easily seen from the following argument: Returning to Eq.~\eqref{eqnA2}, the linear independence of $\{I_B,\hat\BC_1,\hat\BC_4\}$ implies that for some set of coefficients, $a_j,~j=1,2,3$,
\begin{align}\label{eqnA4}
	a_0\AC &= c_7I_A + \left(\frac{1}{6}c_5 - c_7\right)\left(\hat\AC_1 + \hat\AC_3\right),\notag\\
	a_1\AC &= \left(c_6 - c_7\right) I_A + \left(c_1 - \frac{1}{3}c_5 - c_6 + c_7\right)\hat\AC_1 + \left(\frac{1}{2}c_2 - c_6\right)\hat\AC_2 + \left(\frac{1}{3}c_3 - \frac{1}{3}c_5 + c_7\right)\hat\AC_3,\\
	a_4\AC &= \left(c_6 - c_7\right)I_A + \left(\frac{1}{2}c_4 - \frac{1}{6}c_5 - c_6 + c_7\right)\hat\AC_1 + \left(\frac{1}{2}c_4 - c_6\right)\hat\AC_2 + \left(c_7 - \frac{1}{6}c_5\right)\hat\AC_3.\notag
\end{align}
Now, if Bob measures first, then $\AC\propto I_A$, which tells us that  $c_5=6c_7$, $c_1=c_6+c_7$, $c_2=2c_6=c_4$, and $c_3=3c_7$. This leaves a freedom in choosing, say, $c_6$ and $c_7$, as indicated in Eq.~\eqref{eqnA301}. Though this method may appear easier, it may nonetheless be useful for computational implementations to work directly with $\mathbf{Q}^{all}$.]

We now ask what measurements Alice can make to follow such outcomes in Bob's first measurement. We wish to choose a new basis $\SC^\prime_B$ whose dual basis has only one operator not orthogonal to $\BC$ of Eq.~\eqref{eqnA5}. One way to do this using $\mathbf{Q}^{all}$ is as follows. Define $\omega=e^{2\pi i/3}$, and start with the (non-normalized) Fourier matrix
\begin{align}\label{eqnA501}
\mathbf{F}_3 = \left[\begin{array}{ccc} 1 & 1 & 1 \\
							1 & \omega & \omega^2 \\
							1 & \omega^2 & \omega \end{array}\right].
\end{align}
We may replace the first row of ones in $\mathbf{F}_3$ by zeros (since that first row will lead to an operator that is not orthogonal to $\BC$ in our chosen set of $\tilde\QC_l$ operators), but it is just as well to remove that row entirely, calling the resulting matrix $\mathbf{F}_3^\prime$. Multiply $\mathbf{F}^\prime_3$ on the right by
\begin{align}\label{eqnA6}
\mathbf{M} = \left[\begin{array}{ccc} c_6 - c_7 & 0 & 0 \\
							0 & c_7 & 0 \\
							0 & 0 & c_7 \end{array}\right],
\end{align}
and for the moment assume $c_7\ne0$. The reason this works can be seen in a way similar to the discussion at the end of the paragraph following Eq.~\eqref{eqnA3}. That is, if we multiply the vector of operators, $\left(\tilde\BC_0,\tilde\BC_1,\tilde\BC_4\right)^T$ by $\mathbf{F}_3^\prime \mathbf{M}$, we get a vector of operators that is orthogonal to $\left(c_7\BC_0,\left(c_6-c_7\right)\BC_1,\left(c_6-c_7\right)\BC_4\right)^T$, where the latter is representative of $\BC$ in Eq.~\eqref{eqnA5}. We cannot multiply $\mathbf{Q}^{all}$ by $\mathbf{F}_3^\prime\mathbf{M}$, since the inner dimensions don't match, but what we really need to do is this: tensor $\mathbf{F}_3^\prime\mathbf{M}$ with the $4\times4$ identity matrix $I_4$ to get $I_4\otimes\mathbf{F}_3^\prime \mathbf{M}$ (it is $I_n$ with $n=4$ because the dimension of the span of $\SC_A$ is $4$), and multiply this by $\mathbf{Q}^{all}$ on the right. The result is
\begin{align}\label{eqnA7}
\mathbf{Q}_1^\prime := \left(I_4\otimes\mathbf{F}_3^\prime \mathbf{M}\right)\mathbf{Q}^{all} =\left[\begin{array}{ccccccc}
 0 &   0 &  0  &  0 & 0  & -c_7 & c_6\\
 0 &   0 &  0  &  0 & 0 & -c_7 & c_6\\
 c_7\omega & 0 & 0 &  c_7\omega^2/2 & (c_6-c_7\omega)/6 & c_7 & -c_6\\
 c_7\omega^2 & 0 & 0 &  c_7\omega/2 & (c_6-c_7\omega^2)/6 & c_7 & -c_6\\
 0 & c_7\omega/2 & 0 &  c_7\omega^2/2 & 0 & c_7 &  0\\
 0 & c_7\omega^2/2 & 0 &  c_7\omega/2 & 0 & c_7 &  0\\
 0 & 0 & c_7\omega/3 & 0 & (c_6-c_7\omega)/6 & 0 & -c_6\\
 0 & 0 & c_7\omega^2/3 & 0 & (c_6-c_7\omega^2)/6 & 0 & -c_6\end{array}\right],
\end{align}
We wish to find vectors, $\vec c_1^{\,\prime} = \left(c_1^\prime,\cdots,c_7^\prime\right)^T$ such that $\mathbf{Q}_1^\prime\vec c^{\,\prime} = \vec 0$. Note that the last column of $\mathbf{Q}_1^\prime$ is proportional to $c_6$, the second-to-last one, to $c_7$. If both these quantities are non-zero, the nullspace of this matrix is one-dimensional, corresponding to $I_A\otimes\BC$, or in other words, to the result of Bob's initial measurement. Therefore, Alice cannot measure next unless $c_6=0$ or $c_7=0$. Furthermore, since $I_A\otimes\BC$ is not an outcome of the overall desired measurement, then since we are interested in implementing that measurement without error, Bob's initial measurement must correspond to one of these two choices, $c_6=0$ or $c_7=0$. In the first case, $\BC=c_7\hat\BC_7$, $c_7^\prime$ is unconstrained, and we find that the nullspace is two-dimensional, containing all vectors of the form $\vec c_1^{\,\prime}=\left(c_1^\prime,0,3c_1^\prime,0,6c_1^\prime,0,c_7^\prime\right)^T$. This tells us that Alice can measure next with outcomes of the form $\AC_1\otimes\hat\BC_7$, where
\begin{align}\label{eqnA8}
\AC_1 = c_7^\prime I_A + \left(c_1^\prime -c_7^\prime\right)\left(\hat\AC_1 + \hat\AC_3\right).
\end{align}
 
On the other hand, if $c_7=0$ and $\BC = c_6\hat\BC_6=c_6\left(\hat\BC_1+\hat\BC_4\right)$, our choice of $\mathbf{F}_3^\prime \mathbf{M}$ above corresponds to operators $\tilde\QC_l$ all of the form $\AC\otimes\tilde\BC_0$ for some $\AC$, in particular excluding operators of the form $\AC^\prime\otimes\left(\tilde\BC_1-\tilde\BC_4\right)$, which should be included since $\left(\tilde\BC_1-\tilde\BC_4\right)$ is orthogonal to $\BC$. So we need to start over by instead choosing
\begin{align}\label{eqnA9}
\mathbf{M}^\prime = \left[\begin{array}{ccc} 1 & 0 & 0 \\
							0 & 1 & -1\end{array}\right],
\end{align}
and multiplying $\mathbf{Q}^{all}$ by $I_4\otimes\mathbf{M}^\prime$ to obtain
\begin{align}\label{eqnA10}
\mathbf{Q}_2^\prime := \left(I_4\otimes\mathbf{M}^\prime\right)\mathbf{Q}^{all} =\left[\begin{array}{ccccccc}
 0 &   0 &  0  &  0 & 0  & 0 & 1\\
 0 &   0 &  0  &  0 & 0 & 0 & 0\\
 0 & 0 & 0 &  0 & 1/6 & 0 & -1\\
1 & 0 & 0 &  -1/2 & -1/6 & 0 & 0\\
 0 &   0 &  0  &  0 & 0 & 0 & 0\\
 0 & 1/2 & 0 &  -1/2 & 0 & 0 &  0\\
 0 & 0 & 0 & 0 & 1/6 & 0 & -1\\
 0 & 0 & 1/3 & 0 & -1/6 & 0 & 0\end{array}\right].
\end{align}
In this case, we seek vectors $\vec c_2^{\,\prime} = \left(c_1^\prime,\cdots,c_7^\prime\right)^T$ such that $\mathbf{Q}_2^\prime\vec c^{\,\prime} = \vec 0$. Now the second-to-last column of $\mathbf{Q}_2^\prime$ is all zeroes, so $c_6^\prime$ is unconstrained. We find that the nullspace of $\mathbf{Q}_2^\prime$ is also two-dimensional, containing all vectors of the form $\vec c_2^{\,\prime}=\left(c_1^\prime,2c_1^\prime,0,2c_1^\prime,0,c_6^\prime,0\right)^T$. This tells us that Alice can measure next with outcomes of the form $\AC_2\otimes\hat\BC_6$, where
\begin{align}\label{eqnA1001}
\AC_2 = c_6^\prime I_A + \left(c_1^\prime -c_6^\prime\right)\left(\hat\AC_1 + \hat\AC_2\right).
\end{align}
This completes the analysis of the second round of measurements. To summarize to this point in the protocol, we see that Bob's first measurement can have two distinct outcomes, one being $c_7I_A\otimes\hat\BC_7$ followed by Alice's outcomes of the form given in Eq.~\eqref{eqnA8}, and the other being $c_6I_A\otimes\hat\BC_6$ followed by Alice's outcomes of the form given in Eq.~\eqref{eqnA1001}. Note that since the outcomes of Bob's first measurement must sum to $I_A\otimes I_B$, we have that $c_6=1=c_7$. We next consider what measurements Bob can make following each of these outcomes.

For the third round, following Bob's initial outcome of $\hat\BC_7$ and Alice's subsequent measurement, we need to choose operators $\tilde\QC_l$ orthogonal to $\AC_1$ in Eq.~\eqref{eqnA8}. To do so, we multiply $\mathbf{Q}^{all}$ on the left by $\mathbf{M}_{71}\otimes I_3$, where
\begin{align}\label{eqnA11}
\mathbf{M}_{71} = \left[\begin{array}{cccc} 1 & \omega & 0 & \omega^2 \\
							1 & \omega^2 & 0 & \omega \\
							0 & 0 & 1 & 0 \end{array}\right]
\left[\begin{array}{cccc} c_1^\prime - c_7^\prime & 0 & 0 & 0 \\
							0 & c_7^\prime & 0 & 0 \\
							0 & 0 & 1 & 0 \\
							0 & 0 & 0 & c_7^\prime \end{array}\right]
= \left[\begin{array}{cccc} c_1^\prime - c_7^\prime & \omega c_7^\prime & 0 & \omega^2c_7^\prime \\
							c_1^\prime - c_7^\prime & \omega^2c_7^\prime & 0 & \omega c_7^\prime \\
							0 & 0 & 1 & 0 \end{array}\right],
\end{align}
to obtain
\begin{align}\label{eqnA12}
\mathbf{Q}_{71}^{\prime\prime} := \left(\mathbf{M}_{71}\otimes I_3\right)\mathbf{Q}^{all}
	=\left[\begin{array}{ccccccc}
     0 &   0 &         0 &         0 &  -c_7^\prime/6 &           0 & c_1^\prime\\
   \omega c_7^\prime &   0 & \omega^2c_7^\prime/3 &         0 & c_7^\prime/3 & c_1^\prime + \omega^2c_7^\prime &   -c_1^\prime\\
     0 &   0 &         0 &   \omega c_7^\prime/2 & c_7^\prime/6 &  c_1^\prime + \omega^2c_7^\prime &   -c_1^\prime\\
     0 &   0 &         0 &         0 &   -c_7^\prime/6 &  0 & c_1^\prime\\
 \omega^2c_7^\prime &   0 &   \omega c_7^\prime/3 &  0 & c_7^\prime/3 & c_1^\prime + \omega c_7^\prime &   -c_1^\prime\\
     0 &   0 &         0 & \omega^2c_7^\prime/2 & c_7^\prime/6 & c_1^\prime + \omega c_7^\prime &   -c_1^\prime\\
     0 &   0 &         0 &         0 &                     0 &           0 &                 0\\
     0 & 1/2 &         0 &         0 &                     0 &          -1 &                 0\\
     0 &   0 &         0 &       1/2 &                     0 &          -1 &                 0\end{array}\right].
\end{align}
If both $c_1^\prime$ and $c_7^\prime$ are nonzero, the nullspace of this matrix, which we will denote as $\vec c^{\,\prime\prime}$, is one-dimensional, in which case Bob cannot measure next. This would leave a final outcome of the protocol as $\AC_1\otimes\hat\BC_7$, with $\AC_1\not\propto\hat\AC_7$. Therefore, these outcomes must be excluded, since they introduce errors into the protocol. If $c_1^\prime=0$, the nullspace is still one-dimensional and of the form $\left(0,0,0,0,0,0,c_7^{\prime\prime}\right)^T$, so Bob cannot measure following such an outcome. Nonetheless, since this outcome corresponds to $c_7^\prime\hat\AC_7\otimes\hat\BC_7$, it is a valid terminal outcome at the previous round.

On the other hand if $c_7^\prime=0$, then $\mathbf{M}_{71}$ does not provide enough operators $\tilde\QC_l$ to span the space orthogonal to $\AC_1=c_1^\prime\left(\hat\AC_1 + \hat\AC_3\right)$, so instead use
\begin{align}\label{eqnA13}
\mathbf{M}_{72} = \left[\begin{array}{cccc} 1 & 0 & 0 & 0\\
							0 & 1 & 0 & -1 \\
							0 & 0 & 1 & 0 \end{array}\right],
\end{align}
where the first row corresponds to $\tilde\AC_0$, the second row to $\tilde\AC_1 - \tilde\AC_3$, and the third row to $\tilde\AC_2$. Then, we have
\begin{align}\label{eqnA14}
\mathbf{Q}_{72}^{\prime\prime} := \left(\mathbf{M}_{72}\otimes I_3\right)\mathbf{Q}^{all}
	=\left[\begin{array}{ccccccc}
     0 &   0 &         0 &         0 &  0 &           0 & 1\\
   0 &   0 & 0 &         0 &0 & 1 &   -1\\
   0 &   0 & 0 &         0 &0 & 1 &   -1\\
     0 &   0 &         0 &         0 &                     0 &           0 &                 0\\
   1 &   0 &   -1/3 &  0 & 0 & -1 &   0\\
     0 &   0 &         0 & 1/2 &0 & -1 & 0\\
     0 &   0 &         0 &         0 &                     0 &           0 &                 0\\
     0 & 1/2 &         0 &         0 &                     0 &          -1 &                 0\\
     0 &   0 &         0 &       1/2 &                     0 &          -1 &                 0\end{array}\right].
\end{align}
The nullspace of this matrix is of the form $\left(c_1^{\prime\prime},0,3c_1^{\prime\prime},0,c_5^{\prime\prime},0,0\right)$, which is two-dimensional, parametrized by $c_1^{\prime\prime}$ and $c_5^{\prime\prime}$. This corresponds to outcome
\begin{align}\label{eqnA15}
\AC_1\otimes\BC_1 = \hat\AC_5\otimes \left[3c_1^{\prime\prime}\hat\BC_1 + c_5^{\prime\prime}\left(I_B - 2\hat\BC_1 - \hat\BC_4\right)/2\right].
\end{align}
Therefore, Alice's preceding measurement has the two outcomes $3c_1^{\prime}\hat\AC_5\otimes\hat\BC_7=c_1^{\prime}\left(\hat\AC_1+\hat\AC_3\right)\otimes\hat\BC_7$ and $c_7^{\prime}\hat\AC_7\otimes\hat\BC_7=c_7^{\prime}\left(I_A-\hat\AC_1-\hat\AC_3\right)\otimes\hat\BC_7$, and since these must add to $I_A\otimes\hat\BC_7$, their coefficients must be $c_1^{\prime}=1=c_7^{\prime}$. The latter outcome is terminal as $\hat\AC_7\otimes\hat\BC_7$, and the other is followed by Bob's measurement leading to outcomes of the form given in Eq.~\eqref{eqnA15}.

To see what measurements Alice can make after this latter outcome, we must choose the $\tilde\QC_l$ to be orthogonal to its $B$ part, $\BC_1$. This can be done by using
\begin{align}\label{eqnA16}
\mathbf{M}_{73} = \left[\begin{array}{ccc} 3c_1^{\prime\prime} - c_5^{\prime\prime} & \omega c_5^{\prime\prime}/2 & -\omega^2\left(3c_1^{\prime\prime}-c_5^{\prime\prime}\right) \\
							3c_1^{\prime\prime} - c_5^{\prime\prime} & \omega^2c_5^{\prime\prime}/2 & -\omega\left(3c_1^{\prime\prime}-c_5^{\prime\prime}\right) \end{array}\right],
\end{align}
in the form of
\begin{align}\label{eqnA17}
\mathbf{Q}_{73}^{\prime\prime\prime} &:= \left(I_4\otimes\mathbf{M}_{73}\right)\mathbf{Q}^{all}\notag\\
	&=\left[\begin{array}{ccccccc}
0 & 0 & 0 & 0 & 0 & c_5^{\prime\prime}\omega/2 -\left(3c_1^{\prime\prime}-c_5^{\prime\prime}\right)\omega^2 & -\left(3c_1^{\prime\prime}-c_5^{\prime\prime}/2\right)\omega/2\\
0 & 0 & 0 & 0 & 0 & c_5^{\prime\prime}\omega^2/2 -\left(3c_1^{\prime\prime}-c_5^{\prime\prime}\right)\omega &  -\left(3c_1^{\prime\prime}-c_5^{\prime\prime}/2\right)\omega^2\\
c_5^{\prime\prime}\omega/2 & 0 & 0 & -\left(3c_1^{\prime\prime}-c_5^{\prime\prime}\right)\omega^2/2 & -c_1^{\prime\prime}\omega/2 & - c_5^{\prime\prime}\omega/2+\left(3c_1^{\prime\prime}-c_5^{\prime\prime}\right)\omega^2 &  -\left(3c_1^{\prime\prime}-c_5^{\prime\prime}/2\right)\omega/2\\
c_5^{\prime\prime}\omega^2/2 & 0 & 0 & -\left(3c_1^{\prime\prime}-c_5^{\prime\prime}\right)\omega/2 & -c_1^{\prime\prime}\omega^2/2 & -c_5^{\prime\prime}\omega^2/2 +\left(3c_1^{\prime\prime}-c_5^{\prime\prime}\right)\omega & -\left(3c_1^{\prime\prime}-c_5^{\prime\prime}/2\right)\omega^2/2\\
0 &  c_5^{\prime\prime}\omega/4 & 0 & -\left(3c_1^{\prime\prime}-c_5^{\prime\prime}\right)\omega^2/2 & 0 & - c_5^{\prime\prime}\omega/2 +\left(3c_1^{\prime\prime}-c_5^{\prime\prime}\right)\omega^2 &       0\\
0 & c_5^{\prime\prime}\omega^2/4 & 0 & -\left(3c_1^{\prime\prime}-c_5^{\prime\prime}\right)\omega/2 & 0 & -c_5^{\prime\prime}\omega^2/2 +\left(3c_1^{\prime\prime}-c_5^{\prime\prime}\right)\omega &       0\\
0 & 0 &  c_5^{\prime\prime}\omega/6& 0 & -c_1^{\prime\prime}\omega/2 &   0 & \left(3c_1^{\prime\prime}-c_5^{\prime\prime}/2\right)\omega/2\\
0 & 0 & c_5^{\prime\prime}\omega^2/6 & 0 &  -c_1^{\prime\prime}\omega^2/2 &   0 & \left(3c_1^{\prime\prime}-c_5^{\prime\prime}/2\right)\omega^2/2
\end{array}\right].
\end{align}
Except as noted below, the nullspace of this matrix is one-dimensional implying Alice cannot measure next, including when $c_1^{\prime\prime}=0$, which represents the case of the preceding outcome being $c_5^{\prime\prime}\hat\AC_5\otimes\hat\BC_5$. This is therefore an acceptable, terminal outcome of the protocol. Note that the nullspace of $\mathbf{Q}_1^{\prime\prime\prime}$ is not one-dimensional when $c_5^{\prime\prime}=3c_1^{\prime\prime}$, but in that case $\mathbf{M}_{73}$ corresponds to $\tilde\QC_l$ whose $B$-parts are all $\tilde\BC_1$, while we should include $\tilde\BC_0+\tilde\BC_4$ as well, since this is also orthogonal to $\BC_1$ of Eq.~\eqref{eqnA15}. Therefore, we must replace $\mathbf{M}_{73}$ by
\begin{align}\label{eqnA31}
\mathbf{M}_{74} = \left[\begin{array}{ccc} 1 & 0 & 1\\
									0 & 1 & 0 \end{array}\right],
\end{align}
which leads to $ \left(I_4\otimes\mathbf{M}_{74}\right)\mathbf{Q}^{all}$ having a one-dimensional nullspace that corresponds to an outcome proportional to $\left(\hat\AC_1+\hat\AC_3\right)\otimes\left(I_B-\hat\BC_4\right)$. Since the nullspace is one-dimensional this outcome must be terminal, but since it is not one of our desired measurement outcomes, it must be excluded.

The other exception to $\mathbf{Q}_{73}^{\prime\prime\prime}$ having a one-dimensional nullspace is when $c_5^{\prime\prime}=0$, corresponding to $\AC_1\otimes\BC_1=3c_1^{\prime\prime}\hat\AC_5\otimes\hat\BC_1$. This was then one outcome of Bob's preceding measurement, the only other one having been $c_5^{\prime\prime}\hat\AC_5\otimes\hat\BC_5$, which was terminal as we just saw in the preceding paragraph. Since these outcomes must sum to the outcome prior to this measurement, which was $3\hat\AC_5\otimes\hat\BC_7$ (see below Eq.~\eqref{eqnA15}), we have $c_1^{\prime\prime}=1$ and $c_5^{\prime\prime}=6$. Returning to the current measurement outcome by Alice, in this case the nullspace of $\mathbf{Q}_{73}^{\prime\prime\prime}$ is three-dimensional, including all vectors of the form $\left(c_1^{\prime\prime\prime},c_2^{\prime\prime\prime},c_3^{\prime\prime\prime},0,0,0,0\right)$ and allowing outcomes of Alice's subsequent measurement to be any linear combination of $\hat\AC_1,\hat\AC_2,$ and $\hat\AC_3$. However, since the sum of all these outcomes must be equal to the previous outcome, $3\hat\AC_5\otimes\hat\BC_1=\left(\hat\AC_1+\hat\AC_3\right)\otimes\hat\BC_1$ (otherwise, she did not make a complete measurement), and recalling that all of these coefficients $c_j^{\prime\prime\prime}$ are non-negative, we must exclude $\hat\AC_2$ from this list. Furthermore by a similar argument, no subsequent measurements Bob makes can \emph{ever} yield any of the $\hat\BC_j$ other than $\hat\BC_1$. This implies Bob cannot measure again along this branch of the protocol, so Alice will do well to just measure with two outcomes, one being $c_1^{\prime\prime\prime}\hat\AC_1\otimes\hat\BC_1$ and the other being $c_3^{\prime\prime\prime}\hat\AC_3\otimes\hat\BC_1$, and be done with it. Finally, since these outcomes must add to $3\hat\AC_5\otimes\hat\BC_1$, their coefficients must be $c_1^{\prime\prime\prime}=1=c_3^{\prime\prime\prime}$. Thus, we have completed this branch of the protocol, which begins with outcome $I_A\otimes\hat\BC_7$ of Bob's initial measurement, followed by a series of two-outcome measurements, in order as follows: $3\hat\AC_5\otimes\hat\BC_7,\hat\AC_7\otimes\hat\BC_7$, the latter being terminal while the former is followed by $3\hat\AC_5\otimes\hat\BC_1,6\hat\AC_5\otimes\hat\BC_5$, the latter again being terminal and the former followed by $\hat\AC_1\otimes\hat\BC_1,\hat\AC_3\otimes\hat\BC_1$, both of these then being terminal, completing this branch.

This leaves us needing only to analyze the $\hat\BC_6$ branch, that descended from the other of Bob's initial outcomes. We have already seen that Alice's subsequent measurement outcomes must be of the form $\AC_2$ given in Eq.~\eqref{eqnA1001}, and we must find operators orthogonal to these. This can be done by choosing
\begin{align}\label{eqnA18}
\mathbf{M}_{61}= \left[\begin{array}{cccc} c_1^\prime - c_6^\prime & \omega c_6^\prime & \omega^2c_6^\prime & 0\\
							c_1^\prime - c_6^\prime & \omega^2c_6^\prime & \omega c_6^\prime & 0 \\
							0 & 0 & 0 & 1 \end{array}\right],
\end{align}
and then
\begin{align}\label{eqnA19}
\mathbf{Q}_{61}^{\prime\prime} := \left(\mathbf{M}_{61}\otimes I_3\right)\mathbf{Q}^{all}
	=\left[\begin{array}{ccccccc}
     0 &   0 &         0 &         0 &  1/6 &           0 & -1\\
   0 &   0 & 1/3 &         0 & -1/3 & 0 &   1\\
     0 &   0 &         0 &         0 &  -1/6 &           0 & 1\\
     0 &   0 &         0 &         0 &   c_6^\prime\omega/6 &  0 & c_1^\prime + c_6^\prime\omega^2\\
  c_6^\prime\omega &  c_6^\prime\omega^2 /2  & 0 &  0 & -c_6^\prime\omega/3 & c_1^\prime &   -c_1^\prime - c_6^\prime\omega^2\\
     0 &   0 &         0 & -c_6^\prime/2 & -c_6^\prime\omega/6 & c_1^\prime &   -c_1^\prime -  c_6^\prime\omega^2\\
     0 &   0 &         0 &         0 &                c_6^\prime\omega^2/6       &           0 &  c_1^\prime + c_6^\prime\omega  \\
     c_6^\prime\omega^2  & c_6^\prime\omega/2 &         0 &         0 &      -c_6^\prime\omega^2/3  &   c_1^\prime    &  -c_1^\prime - c_6^\prime\omega   \\
     0 &   0 &         0 &    -c_6^\prime /2 &   -c_6^\prime\omega^2/6      &  c_1^\prime   & -c_1^\prime - c_6^\prime\omega   \end{array}\right].
\end{align}
The nullspace of this matrix is one-dimensional, so Bob cannot measure next, unless $c_6^\prime=0$. When $c_6^\prime\ne0$, Alice's preceding outcome is still acceptable for the desired measurement if $c_1^\prime=0$, and then Alice's outcome is terminal, being $c_6^\prime\hat\AC_6\otimes\hat\BC_6$.

When $c_6^\prime=0$, $\mathbf{M}_{61}$ does not provide enough operators $\tilde\QC_l$ to span the space orthogonal to $\AC_2=c_1^\prime\left(\hat\AC_1 + \hat\AC_2\right)=c_1^\prime\hat\AC_4$, so instead use
\begin{align}\label{eqnA20}
\mathbf{M}_{62} = \left[\begin{array}{cccc} 1 & 0 & 0 & 0\\
							0 & 1 & -1 & 0 \\
							0 & 0 & 0 & 1 \end{array}\right].
\end{align}
Then,
\begin{align}\label{eqnA21}
\mathbf{Q}_{62}^{\prime\prime} := \left(\mathbf{M}_{62}\otimes I_3\right)\mathbf{Q}^{all}
	=\left[\begin{array}{ccccccc}
     0 &   0 &         0 &         0 &  0 &           0 & 1\\
   0 &   0 & 0 &         0 &0 & 1 &   -1\\
   0 &   0 & 0 &         0 &0 & 1 &   -1\\
     0 &   0 &         0 &         0 &     1/6 &           0 &                 -1\\
   1 &   -1/2 &   0 &  0 & -1/3 &   0 & 1\\
     0 &   0 &         0 & 0 & -1/6 & 0 & 1\\
     0 &   0 &         0 & 0 &1/6 & 0 & -1\\
     0 & 0 &         1/3 &         0 &                     -1/3 &                 0 &          1 \\
     0 &   0 &         0 &       0 &                     -1/6 &                 0&          1 \end{array}\right].
\end{align}
The nullspace of this matrix is two-dimensional with vectors of the form, $\vec c_2^{\,\prime\prime}=\left(c_1^{\prime\prime},2c_1^{\prime\prime},0,c_4^{\prime\prime},0,0,0\right)^T$, corresponding to outcomes $\AC_2\otimes\BC_2=\left(\hat\AC_1 + \hat\AC_2\right)\otimes\left(c_1^{\prime\prime}\hat\BC_1 + c_4^{\prime\prime}\hat\BC_4/2\right)=2\hat\AC_4\otimes\left(c_1^{\prime\prime}\hat\BC_1 + c_4^{\prime\prime}\hat\BC_4/2\right)$. Comparing to Eq.~\eqref{eqnA1001}, this implies the preceding measurement outcome by Alice was $c_1^\prime\hat\AC_4\otimes\hat\BC_6$, which along with the other (terminal) outcome $c_6^\prime\hat\AC_6\otimes\hat\BC_6$ found in the preceding paragraph, must add to $I_A\otimes\hat\BC_6$. This tells us that $c_1^\prime=2$ and $c_6^\prime=1$.

Following outcome $2\hat\AC_4\otimes\left(c_1^{\prime\prime}\hat\BC_1 + c_4^{\prime\prime}\hat\BC_4/2\right)$, we find Alice's possible subsequent measurements by considering
\begin{align}\label{eqnA22}
\mathbf{M}_{63}= \left[\begin{array}{ccc} 1 & 0 & 0\\
							0 & c_4^{\prime\prime} & -2c_1^{\prime\prime} \end{array}\right],
\end{align}
and then
\begin{align}\label{eqnA23}
\mathbf{Q}_{63}^{\prime\prime} := \left(I_4\otimes\mathbf{M}_{63}\right)\mathbf{Q}^{all}
	=\left[\begin{array}{ccccccc}
     0 &   0 &         0 &         0 &  0 &           0 & 1\\
   0 &   0 & 0 &         0 & 0 & c_4^{\prime\prime}  -2c_1^{\prime\prime} &    2c_1^{\prime\prime} - c_4^{\prime\prime} \\
     0 &   0 &         0 &         0 &  -1/6 &           0 & 1\\
     c_4^{\prime\prime} &   0 &         0 &   -c_1^{\prime\prime}  & \left( c_1^{\prime\prime} - c_4^{\prime\prime}\right)/3 &  2c_1^{\prime\prime} - c_4^{\prime\prime} & c_4^{\prime\prime}  -2c_1^{\prime\prime}\\
     0 &   0 &         0 &         0 &  0 &           0 & 0\\
     0 &   c_4^{\prime\prime}/2 &         0 & -c_1^{\prime\prime}/2 & 0 & 2c_1^{\prime\prime} - c_4^{\prime\prime} &   0\\
     0 &   0 &         0 &         0 &  1/6 &           0 & -1\\
     0 &   0 &         c_4^{\prime\prime}/3 &    0 &   \left( c_1^{\prime\prime} - c_4^{\prime\prime}\right)/3    &  0   & c_4^{\prime\prime}  -2c_1^{\prime\prime}  \end{array}\right].
\end{align}
The nullspace of this matrix is one-dimensional, implying Alice cannot measure next, unless $c_4^{\prime\prime}=0$ or $c_4^{\prime\prime}=2c_1^{\prime\prime}$. Therefore, if $2c_1^{\prime\prime}\ne c_4^{\prime\prime}\ne0$, Bob's preceding measurement outcome had to be terminal. If $c_1^{\prime\prime}\ne0$ so $\BC_2\not\propto\hat\BC_4$, and since $\AC_2\propto\hat\AC_1+\hat\AC_2=\hat\AC_4$, these outcomes must be excluded, as they lead to errors in the final measurement. If $c_1^{\prime\prime}=0$, Alice also cannot measure next, but this yields the valid, terminal measurement outcome, $c_4^{\prime\prime}\hat\AC_4\otimes\hat\BC_4$ from the preceding measurement by Bob. The case of $c_4^{\prime\prime}=2c_1^{\prime\prime}$, corresponding to a preceding outcome of $2c_1^{\prime\prime}\hat\AC_4\otimes\hat\BC_6$, must also be excluded for the following reason. In this case the nullspace is two-dimensional, with vectors of the form $\left(c_1^{\prime\prime\prime},2c_1^{\prime\prime\prime},0,2c_1^{\prime\prime\prime},0,c_6^{\prime\prime\prime},0\right)^T$, corresponding to outcomes $\left[c_6^{\prime\prime\prime}I_A+\left(c_1^{\prime\prime\prime}-c_6^{\prime\prime\prime}\right)\left(\hat\AC_1+\hat\AC_2\right)\right]\otimes\hat\BC_6$. If $c_6^{\prime\prime\prime}=0$, Alice's measurement outcome is $2c_1^{\prime\prime\prime}\hat\AC_4\otimes\hat\BC_6$, so Alice did not actually measure. So we require that for these outcomes, $c_6^{\prime\prime\prime}>0$ (recall that all these $c$ coefficients are non-negative). However, since Alice is making a complete measurement, the sum of all her outcomes must be $\hat\AC_4=\hat\AC_1+\hat\AC_2$, which is impossible if even one of those outcomes has $c_6^{\prime\prime\prime}>0$, which would leave a non-vanishing contribution of $I_A$. Hence, this type of outcome must be excluded.

Finally,  if $c_4^{\prime\prime}=0$, for which the preceding measurement outcome was $c_1^{\prime\prime}\left(\hat\AC_1+\hat\AC_2\right)\otimes\hat\BC_1$, the nullspace is three-dimensional, including all vectors of the form $\left(c_1^{\prime\prime\prime},c_2^{\prime\prime\prime},c_3^{\prime\prime\prime},0,0,0,0\right)$ and allowing outcomes of Alice's subsequent measurement to be any linear combination of $\hat\AC_1,\hat\AC_2,$ and $\hat\AC_3$. However, similarly to what we argued above, we must here exclude outcome $\hat\AC_3$, and Bob may as well just measure with the two terminal outcomes $\hat\AC_1$ and $\hat\AC_2$ to complete this branch of the protocol. Considering the two allowed outcomes of Bob's preceding measurement, $c_4^{\prime\prime}\hat\AC_4\otimes\hat\BC_4$ and $c_1^{\prime\prime}\hat\AC_4\otimes\hat\BC_1$, we must have $c_1^{\prime\prime}=2=c_4^{\prime\prime}$, since these must add to $2\hat\AC_4\otimes\hat\BC_6$, the outcome obtained beforehand.

Thus, we have completed this branch of the protocol, which begins with outcome $I_A\otimes\hat\BC_6$ of Bob's initial measurement, followed by a series of two-outcome measurements, in order as follows: $2\hat\AC_4\otimes\hat\BC_6,\hat\AC_6\otimes\hat\BC_6$, the latter being terminal while the former is followed by $2\hat\AC_4\otimes\hat\BC_1,2\hat\AC_4\otimes\hat\BC_4$, the latter again being terminal and the former followed by $\hat\AC_1\otimes\hat\BC_1,\hat\AC_2\otimes\hat\BC_1$, both of these then being terminal, completing this branch.

\begin{figure}
\includegraphics{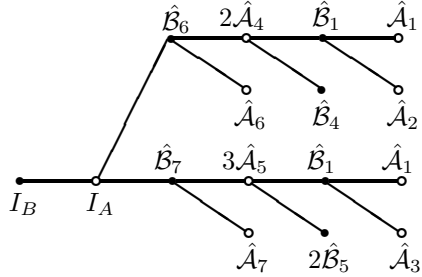}
\caption{\label{figA1}Summary of the results of this appendix, depicting an LOCC protocol for the measurement of Eq.~\eqref{eqnA1}. [Note that only the local operators are indicated next to each node in this diagram, rather than the full multipartite operator.]}
\end{figure}

A summary of these results is depicted as an LOCC tree in Fig.~\ref{figA1}.


%

\end{document}